\documentclass[12pt,preprint]{aastex}
\usepackage{ulem}
\usepackage{url}

\shorttitle{Cosmic Ray removal in fiber spectroscopic image}
\shortauthors{Bai et al.}

\begin{document}

\title{Cosmic Ray removal in fiber spectroscopic image}

\author{Zhongrui Bai\altaffilmark{1,2}, Haotong Zhang\altaffilmark{*}\altaffilmark{2},
Hailong Yuan\altaffilmark{2}, Jeffrey L.  Carlin\altaffilmark{3}, Guangwei Li\altaffilmark{2},
Yajuan Lei\altaffilmark{2}, Yiqiao Dong\altaffilmark{2}, Huiqin Yang\altaffilmark{2},
Yongheng Zhao\altaffilmark{1,2}, Zihuang Cao\altaffilmark{2}}
\email{htzhang@bao.ac.cn}

\altaffiltext{1}{University of Chinese Academy of Sciences, Beijing 100049, China}
\altaffiltext{2}{Key Lab for Optical Astronomy, National Astronomical Observatories, Chinese Academy of Sciences, Beijing 100012, China}
\altaffiltext{3}{LSST and Steward Observatory, 933 North Cherry Avenue, Tucson, AZ 85721, USA}
\altaffiltext{*}{corresponding author}
\begin{abstract}
Single-exposure spectra in large spectral surveys are valuable for time domain studies such as stellar variability, but
there is no available method to eliminate cosmic rays for single-exposure, multi-fiber spectral images. In this paper,
we describe a new method to detect and remove cosmic rays in multi-fiber spectroscopic single exposures.
Through the use of two-dimensional profile fitting and a noise model that considers the position-dependent errors,
we successfully detect as many as 80\% of the cosmic rays and correct the cosmic ray polluted pixels to an average
accuracy of 97.8\%. Multiple tests and comparisons with both simulated data and real LAMOST data show that the
method works properly in detection rate, false detection rate, and validity of cosmic ray correction.
\end{abstract}


\section{Introduction}
Cosmic rays (CRs, hereafter) are high-energy particles that
generate randomly distributed, large signals on charge-coupled
devices (CCDs), which could affect the measured fluxes of
astronomical objects if not detected or removed properly.
Generally, CRs are removed by combining three or more
exposures of the same field \citep{whi94, fre95, fru97, gru14, des16}
, as they are unlikely
to hit the same pixel in more than one exposure. However,
multiple exposures are not always available. Furthermore, there
are certain situations in which CR detection in single exposures
is desired, such as in time domain studies.

Various methods have been developed for identifying and
replacing CRs in CCD data of single exposures, including
median filtering (e.g., Dickinson’s IRAF tasks
\scriptsize QZAP\normalsize{}, \scriptsize XZAP\normalsize{}, and \scriptsize XNZAP\normalsize{}),
applying a threshold on the contrast (e.g., IRAF
task \scriptsize COSMIC-RAYS\normalsize{ }),
trainable classification \citep{mur92, sal95, ber96},
convolution with adapted point-spread functions \citep[PSFs; ][]{rho00},
Laplacian edge detection \citep{van01},
analysis of the flux histogram \citep{pyc04}
and a fuzzy logic-based method \citep{sha05}.
All of the median filtering or PSF methods remove small CRs
from well-sampled data effectively, but problems arise when
CRs affect more than half the size of the filter or when the PSF
is smaller than the filter \citep{van01}.
All of the
methods listed above are designed for photometric data except
those of \citet{van01} and  \citet{pyc04}, which work for long-slit spectroscopic data.

\citet{far05} made a comparison between different methods
including the IRAF script \scriptsize JCRREJ2 \normalsize{} of \citet{rho00},
the IRAF routine L.A.C\scriptsize{OSMIC}\normalsize{} of \citet{van01},
the C script of \citet{pyc04} and the IRAF task \scriptsize XZAP\normalsize{} on photometric images.
In that paper, Farage concluded that L.A.C\scriptsize{OSMIC}\normalsize{} provided the
best performance, with a detection efficiency of 86\% on the real
data sample, whereas other methods could at most detect 78\% of
the CRs. Increasing object density reduces the efficiency of
detection \citep{far05},
which is unfortunately unavoidable in
multi-fiber spectroscopic data where the signals are always
dense. Although L.A.C\scriptsize{OSMIC}\normalsize{} efficiently detects CRs, it replaces
the identified CR candidates with the median value of the
surrounding good pixels \citep{van01},
which is improper
when the CR hits are on the ridge or slope of the profile.

There have been no specific efforts to solve this problem on
multi-fiber spectroscopic data, which present distinct challenges 
compared to photometric data. Multi-fiber images do not
have clear isolated point or extended sources as in the
photometric data, and the long stripe-like multi-fiber spectra
occupy large contiguous regions so that the available area for
the local “background” is much smaller than in the photometric
data. Methods with median filtering or interpolation of
neighboring pixels are less effective in this case.

We present an algorithm to detect and replace CRs for
Large Sky Area Multi-Object fiber Spectroscopic Telescope
(LAMOST) single-exposure images based on a two-dimen-
sional (2D) profile fitting of the spectral aperture. We first pick
out CR candidates with Laplacian edge detection and construct
a 2D function to fit the image profile in small segments along
the spectral trace with these candidates masked out; the final
CR list is generated by comparing the fitting residual with a
noise model depending on position, and the CR polluted pixels
are replaced with the corresponding value of the 2D function.
This method is applied to the data processing of LAMOST; in
principle, it can also be used for other multi-fiber spectral data
after minor modification.

We describe LAMOST data in \S2. The algorithm is explained in \S3.
In \S4, we give some examples and analyze the properties of the algorithm.
Finally, in \S5 we summarize our work.

\section {LAMOST Data}\label{sec2}

LAMOST \citep{cui12} is a fiber spectroscopic telescope
equipped with 4000 fibers feeding 16 spectrographs. Each
spectrograph, holding 250 fibers, is split into blue
(3700–5900 \AA) and red (5700–9000 \AA) arms by a dichroic
mirror. Groups of 250 spectra are recorded by two $4k{\times}4k$
CCDs at the blue and red end, respectively. The typical
duration for a single LAMOST exposure ranges from 600 to
1800s, depending on the target brightness and weather
conditions. A considerable number of CRs hit the images
during the exposure; for example, in a typical 1800s image, the
number of pixels polluted by CRs is about
$2\times10^4$.

\begin{figure}
\epsscale{.99}\plotone{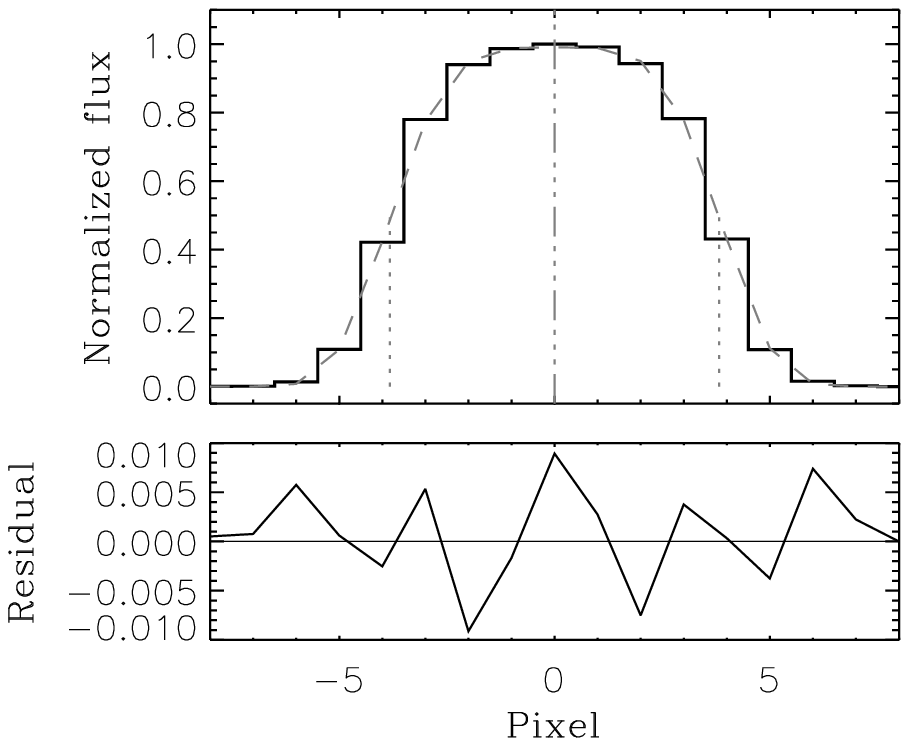}
\caption{Upper panel: the typical shape of a fiber cross section of LAMOST. The black solid line is the data, while the dashed line is the fitted S\'ersic profile. The dot-dashed line shows the center of the profile, and the width between the dotted lines is the FWHM. Lower panel: the residual of the fitting.
\label{fibersec}}
\end{figure}

The size of a LAMOST image, of which the dispersion
direction is along the vertical direction, is 4096$\times$4136 pixels.
In the spatial direction, the typical distance between two
adjacent fibers is 15$\sim$16 pixels. As shown in Figure \ref{fibersec},  the
cross section of the fiber profile in the spatial direction could
be well described by a S\'ersic function \citep{ser68, cle02}:
\begin{equation}\label{eq:1}
P(x)=\alpha{e}^{-\frac{|x-\beta|^\delta}{\delta\gamma^\delta}},
\end{equation}
where $\alpha$, $\beta$, $\gamma$, $\delta$ are parameters to be derived. The typical
full width at half maximum (FWHM) of the profile is about 7$\sim$8 pixels.
If $d=|x-\beta|$ is the distance from a given pixel
to the fiber profile center in the row (or horizontal/spatial)
direction, according to Figure ~\ref{fibersec}, the flux at $d=8$ is less than
0.01\% of those at the profile peak. To avoid fiber to fiber cross
talk, the magnitude range of objects observed in one LAMOST
observation is constrained to be less than 5 magnitudes. In the
extreme case, the contribution from the 5 magnitude brighter
neighbor to the pixel at $d=8$ could be ignored, so the fluxes in
the pixels of $d\leq 8$ could be considered as the flux from the
fiber itself. In 2D data reduction, $d=8$ is chosen as the
aperture for spectrum extraction. The spectral resolution of
LAMOST is about 1800, which corresponds to a FWHM about
5 pixels in the dispersion direction. The PSF changes gradually
with position on the CCD chip, but could be considered as
constant in a small region (e.g., 20 pixels); we will take
advantage of this characteristic to improve the cosmic ray
rejection.

\section {Cosmic Ray Detection and Rejection}

CRs are detected and replaced in three steps. First, we use
Laplacian edge detection \citep{van01} to generate a raw
CR candidate list. Second, for each fiber, pixels within $d=8$
of the fiber trace center are divided into small blocks; each
block is then fitted by a 2D profile with those raw CR
candidates masked out. The final CR list is determined by
comparing the fitting residual with a noise model considering
both the intrinsic noise and the uncertainty introduced by
profile fitting; the pixels polluted by CRs are replaced by the
corresponding fitted value. The details are as follows.

\subsection {Laplacian edge detection \label{sec:LACOS}}
Laplacian edge detection has been widely used for highlighting 
boundaries in processing digital images \citep[e.g., ][]{gon92}.
\citet{van01} was the first to apply the
method to detect CRs in astronomical images. Their publicly
available program, L.A.C\scriptsize{OSMIC}\normalsize{}, successfully detects CRs in
both photometric and long-slit spectroscopic images. We use a
similar method to that in Section 3 of \citet{van01} to
pick out the raw CR candidates. Since further details can be
found in that paper, only basic steps are listed here.

The original image $I$ with the size of 
${n_x}\times{n_y}$ is subsampled into ${2n_x}\times{2n_y}$:
\begin{equation}
I^{(2)}_{i,j}=I_{int[(i+1)/2],int[(j+1)/2]},
\end{equation}
where $i=1,\cdots,2n_x$ and $j=1,\cdots,2n_y$.
The subsampled image is then convolved with a Laplacian kernel:
\begin{equation}
\L=\mathcal{L}*I^{(2)},
\end{equation}
where $\mathcal{L}$ is the Laplace operator, $*$ denotes convolution.
The Laplace operator $\mathcal{L}$ in the above convolution is

\begin{equation}
\mathcal{L}=\frac{1}{4}\left( \begin{array}{ccc}
                                0 & -1 & 0 \\
                                -1& 4 & -1 \\
                                0 & -1 & 0
                                \end{array} \right).\
\end{equation}

Since CRs are positive in $\L$, all the negative values are set to zero.
$\L$ is then resampled to ${n_x}\times{n_y}$:
\begin{equation}
\L^{+}_{i,j}=\frac{1}{4}(\L_{2i-1,2j-1}+\L_{2i-1,2j}+\L_{2i,2j-1}+\L_{2i,2j}),
\end{equation}
where $i=1,\cdots,n_x$ and $j=1,\cdots,n_y$.

 The original image is median filtered with a 
${5}\times{5}$ box to construct the noise model
\begin{equation}
N_{m5}=\frac{1}{g}\sqrt{g(I_{m5})+\sigma_{rd}^2}, \label{nm5}
\end{equation}
where $g$ is the gain in electrons per ADU, $I_{m5}$ is 
the image median filtered  by a ${5}\times{5}$ box,
and $\sigma_{rd}$ is the readout noise in
electrons. The Laplacian image is then divided by the noise
model and the subsampling factor to obtain the deviations from
the expected Poisson fluctuations:
\begin{equation}
S=\frac{\L^{+}}{2N_{m5}}.
\end{equation}
All structures that are smooth on scales of $\geq{5}$ pixels are removed by
a ${5}\times{5}$ median filter:
\begin{equation}
S^{'}=S-S_{m5}.
\end{equation}

All pixels that meet $S^{'}>\sigma_{lim}$ are identified as CR candidates,
where $\sigma_{lim}$ is a given threshold.
We adopt $\sigma_{lim}=4.5$, similar to
L.A.C\scriptsize{OSMIC}\normalsize{}.

\subsection{Fiber profile fitting}

Fiber traces are closely aligned on multi-fiber spectral
images. In contrast to photometric images, useful signals are
quite fully rather than sparsely distributed. As pointed out by
Farage \cite{far05}, the increasing object density will certainly
reduce the efficiency of the methods designed for photometric
data. The ramp on either side of the ridge of the fiber trace is
quite steep, so it is hard for the edge detection method to
discriminate between real CR hits and good pixels on the ramp,
leading to a drop in the detection rate and a rise in the false
detection rate. Furthermore, replacing the CR polluted pixels
with the median of the surrounding pixels is seemingly unsafe.
Along the fiber trace, the shape of the PSF changes slowly. If
the PSF is well sampled, then CR discrimination could be
improved by its shape difference from the PSF.

For an image $I$ of ${n_x}\times{n_y}$ pixels with $n_f$ fibers, the pixels
close to the fiber trace center contribute the most to the
extracted flux. CRs hitting on these areas will introduce large
errors in the final spectra, while those in the trough between 
fibers have much less impact. Consider a small spectral
segment centered on column $[c_{kj}]$ and row $j$,
where $k=1,\cdots,n_f$, $j=1,\cdots,n_y$ and $[c_{kj}]$ is the trace center of
the $k$th fiber at row $j$($[]$ denotes the round off of the quotation).
Since both the trace center and the shape of the PSF change
slowly inside the segment, the shape of the segment could be
fitted with a product of two orthogonal vectors:
\begin{eqnarray}
I_{xy} & = & \mathcal{F}_{xy}+\varepsilon_{xy} \nonumber\\ & = & S(x)P(y)+\varepsilon_{xy} \label{eqfit},
\end{eqnarray}
where $S(x)$ is the fiber profile in the spatial direction, $P(y)$ is a
polynomical to describe the flux variation in the dispersion
direction, and $\varepsilon_{xy}$ is noise. The size of $x$ is the same as the
aperture for flux extraction, which is set to $d=8$ for LAMOST, as discussed in Section \ref{sec2}.
The size of $y$ is chosen to be small enough to keep  $P(y)$  
smooth but larger than the size of single CR hits,
so that $P(y)$ can be fitted with a low-order
polynomial and the CR polluted pixels could be better
estimated by interpolation. For LAMOST, the segment size is
set $17\times 9$, i.e., $x=[c_{kj}]-8,\cdots,[c_{kj}]+8$  and 
$y=j-4,\cdots,j+4$. We do not try to fit the pixels in the
bottom of the valley between fibers, since they contribute little
to the final spectrum.

The shape of $S(x)$ is determined by the output pupil of the
fiber and instrument distortion. Although the typical shape of
 $S(x)$ could be described by a S\'ersic function (Eq. \ref{eq:1}),
the
actual shape deviates occasionally from the analytic function
when the optical distortion is large at the edge of the image or
the coupling between the fiber output pupil and the slit is
imperfect. Due to the above reason, $S(x)$ is constructed with an
empirical profile rather than an analytic function. All the
profiles at $y=j-10\sim{j+10}$ are first normalized, center
justified in sub-pixel scale and then averaged to derive $S(x)$ 
with the CR candidates masked out. Fixing the form of $S(x)$, 
the  polynomial coefficients of $P(y)$  are derived by least-square
surface fitting to the flux in the segment with the CR candidates
masked out. A fitted image $\mathcal{F}$ is generated after all segments are handled.

\subsection {Cosmic ray selection}
A new list of CRs is generated by comparing the noise
model with the residual of the image fitting without reference to
the old CR list. The noise or uncertainty of our method comes
from two parts: one is the intrinsic noise of the input signal, i.e.,
Poisson noise from the object and the readout noise from the
CCD circuit; the other part of the noise comes from the defect
of the profile fitting, which is larger when the profile changes
more dramatically. Basically, the first part is related to time and
the second part is related to position, which could be illustrated as:
 \begin{eqnarray}
    \Delta{\mathcal{F}} & = & |\frac{\partial{\mathcal{F}}}{\partial{t}}\cdot{\Delta{t}}|+\left(|\frac{\partial{\mathcal{F}}}{\partial{x}}\cdot{\Delta{x}}|+|\frac{\partial{\mathcal{F}}}{\partial{y}}\cdot{\Delta{y}|}\right)\nonumber \\
     & = &\frac{1}{g} \sqrt{g\mathcal{F}+\sigma_{rd}^2}+ \left(|\frac{\partial{\mathcal{F}}}{\partial{x}}\cdot{\Delta{x}}|+|\frac{\partial{\mathcal{F}}}{\partial{y}}\cdot{\Delta{y}}|\right).
       \label{newnoise}
 \end{eqnarray}
The first term is sufficient to pick out CRs for most cases
when the fit is good, yet it is necessary to add the position-
dependent term to avoid false detections in regions where the
fit is not perfect. The following steps are implemented to
reject the CRs:
\begin{enumerate}
\item  The position dependent terms in \ref{newnoise} are
calculated by the average gradients at each pixel.
Convolving the fitted image  with the following four
arrays:
\begin{equation}
\begin{array}{cccccc}
A_1 & = & \frac{1}{2} \left(\begin{array}{ccc}
                                0 & 0 & 0 \\
                                -1& 0 & 1 \\
                                0 & 0 & 0
                                \end{array} \right)\  &
A_2 & = & \frac{1}{2} \left( \begin{array}{ccc}
                                0 & 1 & 0 \\
                                0 & 0 & 0 \\
                                0 & -1 & 0
                                \end{array} \right)\ \\
A_3 & = & \frac{1}{2\sqrt{2}}\left( \begin{array}{ccc}
                                -1 & 0 & 0 \\
                                0 & 0 & 0 \\
                                0 & 0 & 1
                                \end{array} \right)\ &
A_4 & = & \frac{1}{2\sqrt{2}}\left( \begin{array}{ccc}
                                0 & 0 & 1 \\
                                0 & 0 & 0 \\
                                -1 & 0 & 0
                                \end{array} \right)\
\end{array},
\end{equation}

a gradient array can be derived as
\begin{equation}
G=\frac{1}{2}(|A_1*\mathcal{F}|+|A_2*\mathcal{F}|+|A_3*\mathcal{F}|+|A_4*\mathcal{F}|).
\end{equation}

\item   Noise models are constructed with and without the
second term in Equation \ref{newnoise}, respectively,
\begin{equation}
N_1=\frac{1}{g}\sqrt{g\mathcal{F}+\sigma_{rd}^2},
\end{equation}
and
\begin{equation}
N_2=\frac{1}{g}\sqrt{g\mathcal{F}+\sigma_{rd}^2}+G.
\end{equation}
The noise-weighted differences between the input image $I$ 
and the fitted image $\mathcal{F}$ are defined accordingly:
\begin{equation}
D_1=\frac{(I-\mathcal{F})}{N_1},
\end{equation}
and
\begin{equation}
D_2=\frac{(I-\mathcal{F})}{N_2}.
\end{equation}

All pixels with $D_1>20$ or $D_2>3$ are marked as CR
candidates in this step. A mask array $M$ is generated with
the CR polluted pixels set to 1.

\item{\label{item3}}
By setting the previous limits, the number of fake CRs is
greatly reduced, but real CRs with a low signal-to-noise
ratio (S/N), most of which are indiscernible from noise,
are blocked as well. Considering the consecutive pixels
occupied by a certain CR hit, the pixels at the edge of the
CR hit are more likely to be rejected due to lower signal,
though they should have higher probability to be real than
the single-pixel event. So a lower limit for those
neighbouring pixels will raise the detection rate. To do
this, all of the neighboring pixels are first added back to
the CR list by convolving the mask array $M$ with
\begin{equation}
B = \left( \begin{array}{ccc}
                                1 & 1 & 1 \\
                                1 & 0 & 1 \\
                                1 & 1 & 1
                                \end{array} \right)\ ,
\end{equation}
for the pixels in the expanded CR list, if the corresponding
 $D_1>2$ or $D_2>2$, then they will be added to a new mask  array $M{'}$.

\item  Assuming there are sufficient CRs masked out, the
residual of the fit $(I-\mathcal{F})$ will represent the actual
difference between the original CR-free image and the
fitted image. The difference can be added back to
compensate for imperfect fitting, and a more accurate fit
will help to raise the CR detection rate as follows.

The difference is derived by median filtering the residual $(I-\mathcal{F})$
with the CR candidates $M{'}$ masked out:
\begin{equation}
D_m=[(I-\mathcal{F})(1-M{'})]_{m3},
\end{equation}
where m3 denotes a $3\times 3$ median filter. And the new noise-weighted residual arrays will be
\begin{equation}
D^{'}_1=\frac{(I-\mathcal{F}-D_m)}{N_1},
\end{equation}
and
\begin{equation}
D^{'}_2=\frac{(I-\mathcal{F}-D_m)}{N_2}.
\end{equation}
All pixels with $D^{'}_1>10$ or $D^{'}_2>3$ or $D_1>20$ or $D_2>3$ are masked as CR candidates in this step.

\item  Rerun to Step \ref{item3} and confirm the final CR candidates.
\end{enumerate}
We do not try to fit the pixels that are either in bad fibers or in
the $d>8$ gaps between fibers because they contribute little to
the final extracted spectra. For those pixels, $\mathcal{F}$ is set to 0 and
the CR candidates are selected by simply requiring the noise-
weighted difference between the original image $I$ and the $5\times{5}$ 
median filtered image $I_{m5}$ to be larger than 3:
\begin{equation}
\frac{I-I_{m5}}{N_{m5}}>3,
\end{equation}
where $N_{m5}$ is the same as in Equation \ref{nm5}.

Combining the above CR candidates, the final CR mask is
generated and the value of each CR polluted pixel is replaced
by the corresponding value in the fitted image $\mathcal{F}$.

\subsection {Additional features}

With the profile fitting method, other bad pixels such as the
inherent damaged pixels could be replaced with a reasonable
value once an initial bad pixel map is known.

On a Dell Precision T5500 (eight 2.0 GHz CPUs), the IDL
implementation with one single-threaded processor requires
about 20 minutes for an image of ${4096}\times{4136}$ pixels. Most of
the time is spent on the 2D profile fitting and the timescales
linearly with the number of fibers and image size in the
dispersion direction. As the current version of our program is
not parallelized, the execution time on the current computer is
equivalent to that of a single-core processor and could be
greatly reduced after software parallelization.

\section {Examples and Application}
In this section, tests with both simulated data and real data
are carried out to illustrate the performance of our method. In
these tests, our primary concerns are the following factors: how
many pixels of real CRs are detected (efficiency), how many
pixels are falsely detected as CRs (false detection rate), and the
accuracy of the CR replacement. A better method should have
higher efficiency, lower false detection rate, and proper
restoration of the pixels polluted by CRs.

\subsection {Artificial Images}\label{arti1}
For multi-fiber spectral observations, to increase the
observation efficiency as well as to avoid fiber to fiber cross
talk, it is usually a good strategy to divide the targets into
different plates according to their brightness so that the S/N of
the targets in the same plate are similar in the same exposure
time. For plates with bright magnitude, the exposure time is
short, so the strength of the sky spectrum is low, but the S/N of
the objects is high. On the contrary, the exposure times for the
faint plates are long, so the sky spectrum is strong and the
object spectrum is weak. Since the efficiency strongly depends
on the brightness contrast between the CR and the target, plates
with different target brightness are simulated to test our method
under different situations.

First, we generate a pure CR image of ${4136}\times{4096}$ pixels
with 20,000 CR hits (approximately 10 times of those in a 30
minutes exposure LAMOST image). The shape of each CR is
set to be an ellipse with the major axis randomly distributed in 
1$\sim$10 pixels and the minor axis ranges from 1 to 3 pixels, by
which almost all kinds of CRs in the real image can be
simulated. The direction of the major axis is randomly
distributed in 0$\sim 360^\circ$ and the intensities are uniformly
distributed between 0$\sim$20000 ADUs. All CRs with flux less
than 5, which is at the readout noise level, are set to 0. In total,
227,451 pixels are polluted by CR hits.

Second, to study the method’s performance with different
target brightness, two CR-free images are generated by
combining three consecutive LAMOST exposures of the same
targets. The first image (IMG600) is combined from three 600
second exposures in which the spectra are dominated by the 
strong smooth continuum from the bright objects and the sky
emission lines are relatively weak. In the second one
(IMG1800), the sky emission lines are more prominent (due
to a longer exposure time of 1800 seconds) and the object
continuum is relatively weak (due the faint magnitude). The
final test images are generated by adding the pure CR image to
the CR-free images.

Both IMG600 and IMG1800 are tested by our method and
the IDL version of
 L.A.C\scriptsize{OSMIC}\normalsize{}, respectively. The results are
summarized in Table 1. For IMG600, the efficiency of our
method is 1.9\% higher (73.8\% versus 71.9\%) and the number
of false detections is 5820, two orders of magnitude lower than
that of L.A.C\scriptsize{OSMIC}\normalsize{}.
For IMG1800, our efficiency is 4.5%
higher(80.9\% vies 76.4\%) and the number of false detections
(16,626) is less than half that of L.A.C\scriptsize{OSMIC}\normalsize{(38,912).}

The efficiency of both methods rises more than 4\% from
IMG600 to IMG1800. The reason is that the efficiency, for
those CRs falling coincidentally into the same pixel with the
object spectrum, decreases with the increasing photo noise,
while in this case, the CRs are the same in both simulations but
the object spectra are much brighter in IMG600 therefore the
noise is larger in IMG600 than in IMG1800. Compared with
our method,  L.A.C\scriptsize{OSMIC}\normalsize{} is prone to mistake the wings of the
bright profiles as the sharp edges of CRs, especially when the
contrast between the background and the profile peak is high,
as in IMG600. In IMG1800, as the object brightness decreases,
the contrast and thus the number of false detections drops. Our
profile fitting method successfully bypasses this sharp edge trap
in IMG600, reducing the huge number of fake detections to a
reasonable level. As shown in Table \ref {table1}, for our method, the
number of false detections doubles from IMG600 to IMG1800.
There are two reasons for this problem. First, as the SNR of the
spectrum becomes lower, more faint pixels are mistaken as
CRs (as can be seen from Figure  \ref{falsedetection}). Second, as the exposure
time increases, the intensity of the sky emission lines increases,
but the intensity of the underlying object spectra decreases (for
the magnitude gets much fainter). In this case, the relative
change at some of the exponential wings of the strong sky
emission lines becomes too dramatic to have a good
polynomial fit; the larger residual induced by the improper
fitting leads to an increase of false detections.

\begin{figure}
\includegraphics[scale=.9]{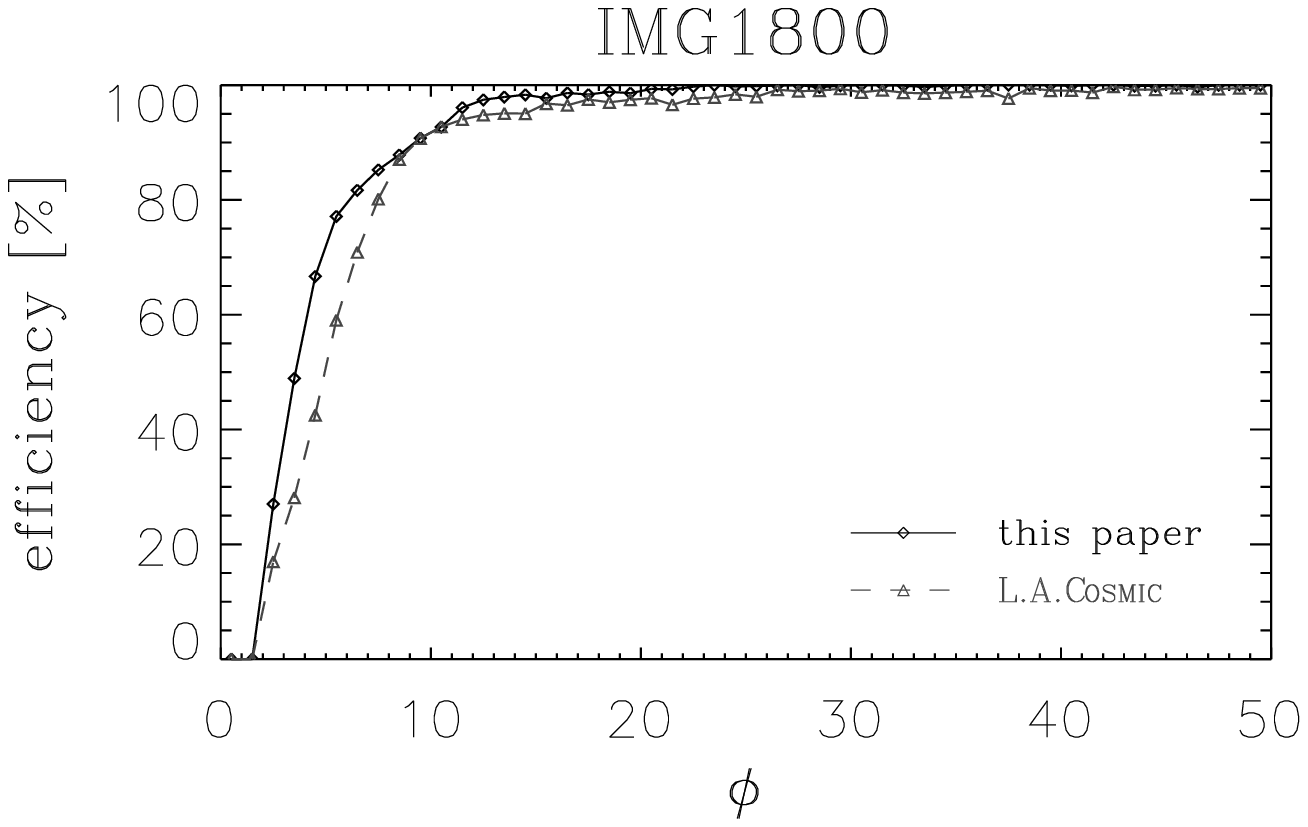}
\includegraphics[scale=.9]{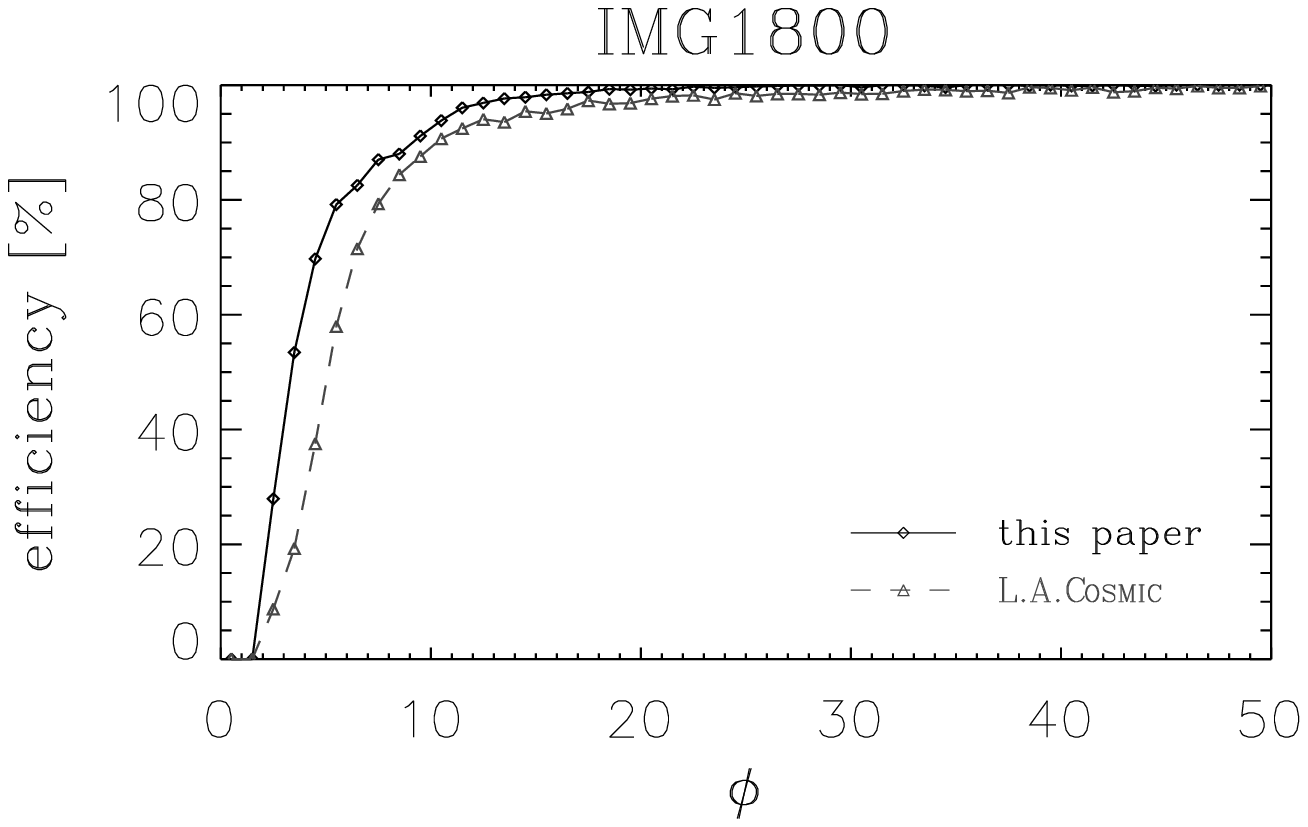}\\
\caption{Efficiency vs. $\phi$ in different simulations. The upper and lower panels are for IMG600 and IMG1800, respectively. The solid and dashed curves are for the method of this paper and L.A.C\scriptsize{OSMIC}\normalsize{}, respectively.
\label{detecteff}}
\end{figure}

If we denote the SNR of a CR polluted pixel as
\begin{equation}
\phi=\frac{f_{CR}}{\sqrt{f_{clean}+\sigma_{rd}^2}},
\end{equation}
where $f_{CR}$ and $f_{clean}$ are the fluxes from  the pure CR and the CR-free image, respectively, $\sigma_{rd}$ is the readout noise,
then Figure  \ref{detecteff} shows the detection efficiency against $\phi$. Most of the
undetected CRs are those with low $\phi$.  The efficiency
remains high for $\phi>10$ then drop quickly when $\phi<10$. The efficiency of our method is higher than that of L.A.C\scriptsize{OSMIC}\normalsize{ }\
in all situations except for $\phi<2$, where the CRs  are too weak to be separated.

\begin{figure}
\epsscale{.99}\plotone{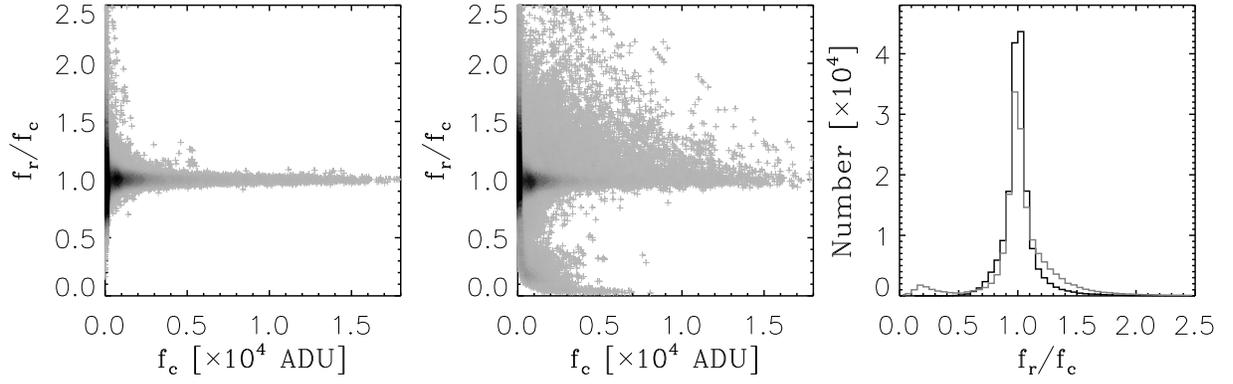}
\caption{Replacement of good detections on IMG600. In this figure,$f_{r}$  is the corrected flux of the CR polluted pixel, while $f_{c}$ is the flux from the corresponding pixel in the CR-free image. The left and middle panels are the replacements using our method and L.A.C\scriptsize{OSMIC}\normalsize{}, respectively; different gray levels represent the relative number density. The right plot shows the histogram of $f_{r}/f_{c}$ in which the black line is our method and the grey line is L.A.C\scriptsize{OSMIC}\normalsize{}.
\label{gooddetection}}
\end{figure}

\begin{figure}
\epsscale{.99}\plotone{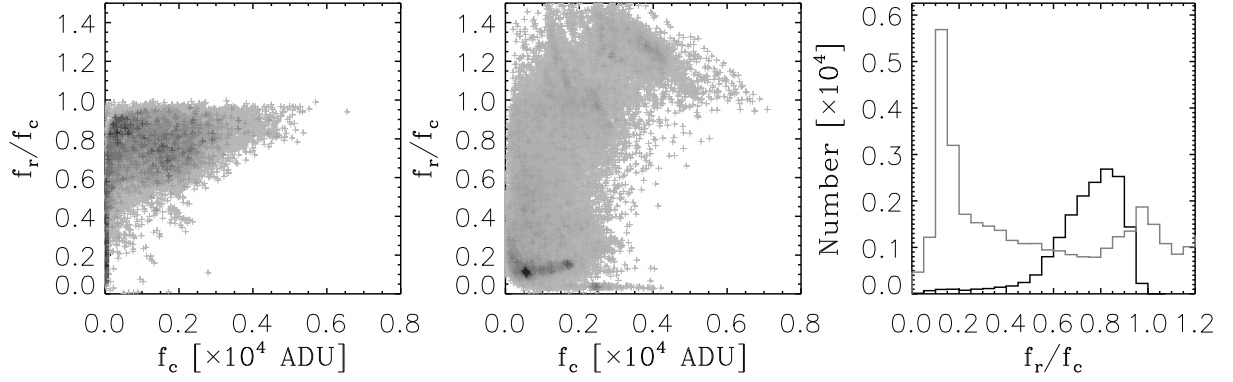}
\caption{Replacement of the false detections on IMG1800. $f_{r}$ and $f_{c}$ are the same as defined in Figure \ref{gooddetection}. The left and the middle panels are the replacements of our method and L.A.C\scriptsize{OSMIC}\normalsize{}, respectively; number density is indicated by different gray levels. The right plot shows the histogram of $f_{r}/f_{c}$, in which the black line is our method and the grey line is L.A.C\scriptsize{OSMIC}\normalsize{}.
\label{falsedetection}}
\end{figure}

In Figure \ref{gooddetection}, the recovered fluxes of IMG600 are compared to
the corresponding fluxes of the CR-free image to see how well
our CR replacement works; as shown in the left and the right
panel, almost all the replacements properly follow the original
fluxes. Also shown in the middle panel is the replacement
performance of L.A.C\scriptsize{OSMIC}\normalsize{};  most of the replacements are
good, but the scatter is larger especially in the high flux region,
which is not unexpected, since its replacement method is not
specially designed for multi-fiber spectra. The performance of
both methods on IMG1800 is similar to IMG600.

\begin{figure}
\epsscale{.9}\plotone{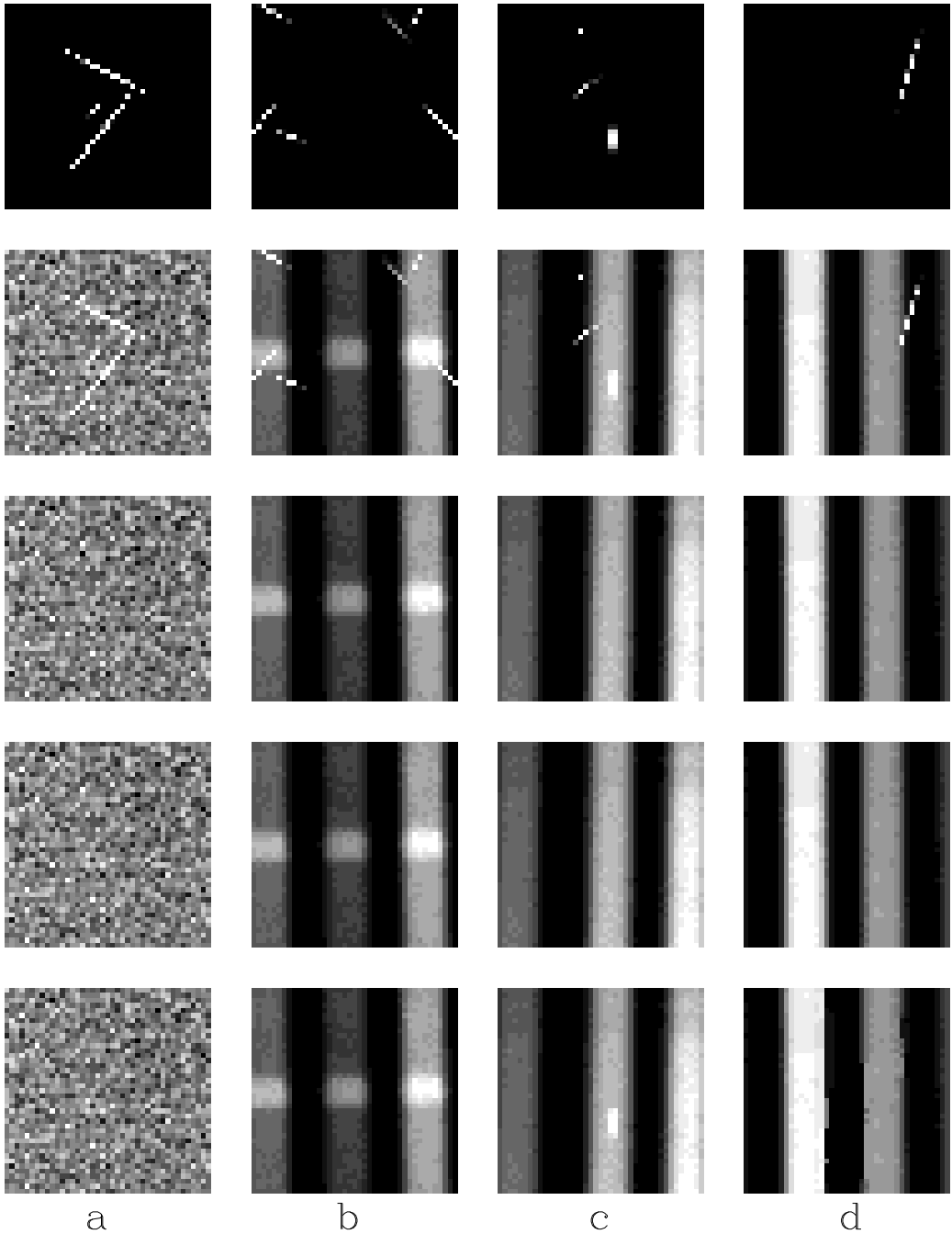}
\caption{Four examples of CR correction. Rows from top to bottom: the pure CR image, the CR added image, the clean image, the result of our method, and the result of L.A.C\scriptsize{OSMIC}\normalsize{}, respectively. Columns $a$ and $b$ show that both methods work well for CRs around low flux fibers. Column $c$ shows that L.A.C\scriptsize{OSMIC}\normalsize{} fails to detect a CR hit on the ridge of a fiber trace. Column $d$ shows that L.A.C\scriptsize{OSMIC}\normalsize{} mistakes too many good pixels near the edge of the bright fiber trace as CR polluted, which is the biggest problem of L.A.C\scriptsize{OSMIC}\normalsize{}.
\label{sam}}
\end{figure}

Since the replacement of the false detections also changes
the flux, causing errors in the spectrum, it’s necessary to test the
replacement on those falsely detected CRs.
 L.A.C\scriptsize{OSMIC}\normalsize{}\ produces too many falsely detected CRs to be
comparable with our method in IMG600, so only the results
of IMG1800 for both methods are shown in Figure  \ref{falsedetection}.
As shown
in the picture, the results of both methods deviate from the true
value. Though our method systematically underestimates the
flux, most of our replaced fluxes concentrate within 80\% of
their true value and the true fluxes of most of the pixels are low,
so the influence on the extracted spectrum should be small.
L.A.C\scriptsize{OSMIC}\normalsize{} results show a large variation, with a large number
of pixels shifting from the true value to very low fluxes. Some
examples are demonstrated in Figure  \ref{sam}.

\begin{figure}
\epsscale{.9}\plotone{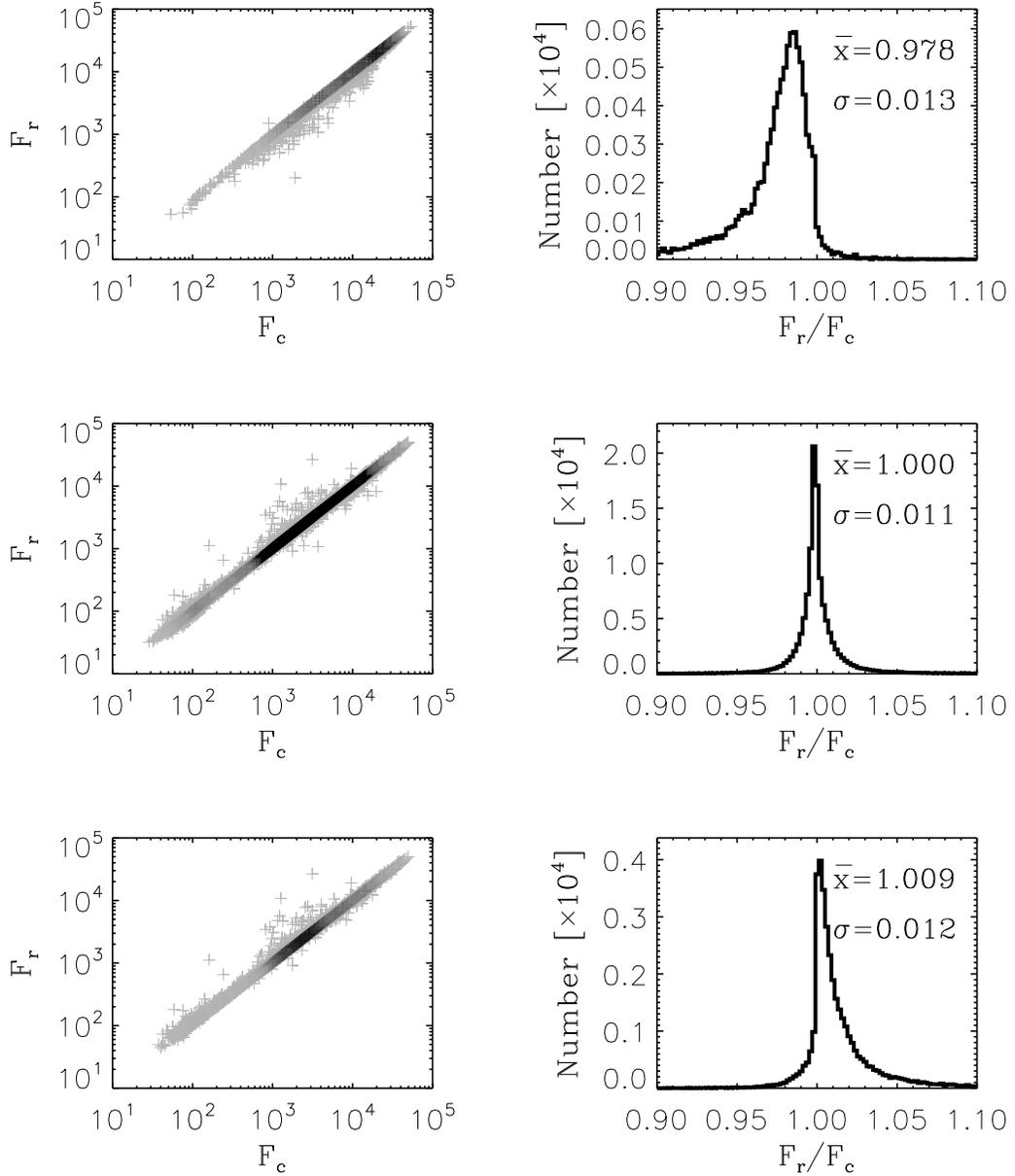}
\caption{Performance of our CR correction. $F_r$ is the spectral flux extracted at the position of CR influences (any detected or fake or undetected CR within the extraction aperture) from the CR corrected image, and $F_r$ is the corresponding flux from the CR-free spectrum. The left column shows the comparison of our result and the clean spectra, with the number density indicated by the grey level. The right column shows the histogram of $F_r/F_c$. Rows from top to bottom show the replacement
performance of falsely detected, correctly detected and undetected CRs, respectively. The biggest deviation (2.2\%) happens in the false detection, since the program tries to replace the falsely detected CRs with lower "correct" values.
\label{extr}}
\end{figure}

\begin{figure}
\epsscale{.99}\plotone{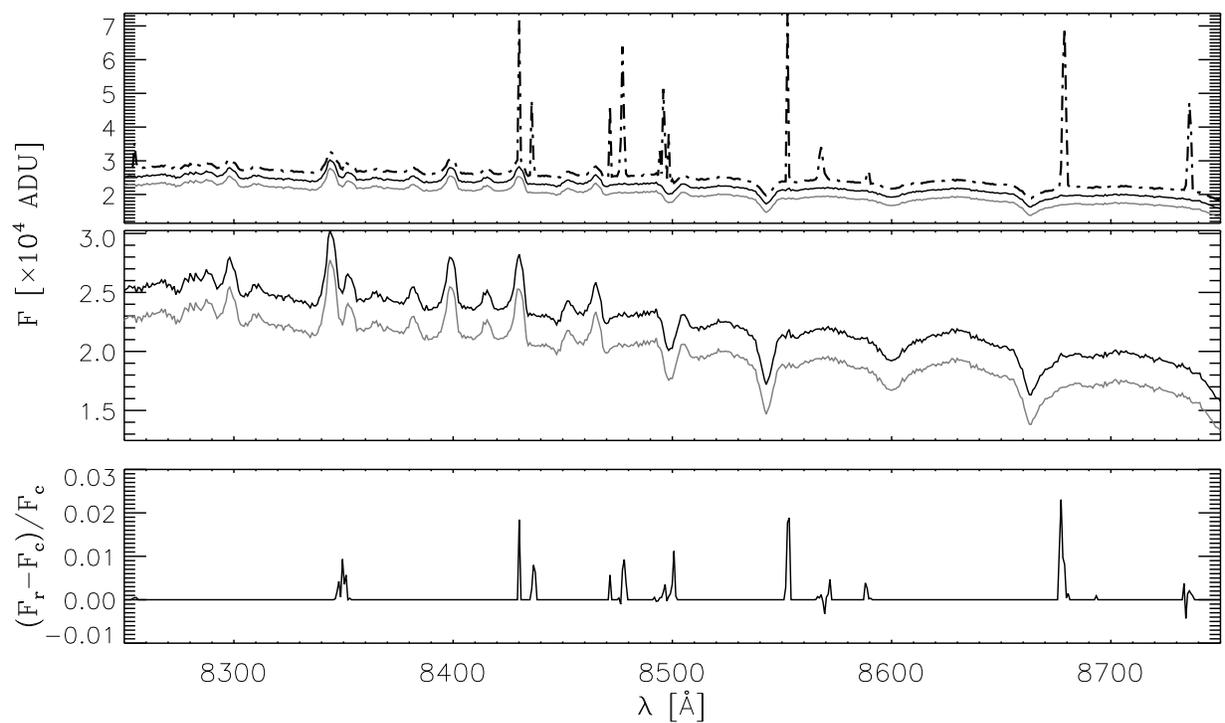}
\caption{Example of extracted spectra. In the top panel, the dotted line is the spectrum extracted from the image without CR correction, the solid black line is the CR- free spectrum and the solid gray line is our CR corrected result. Spectra have been shifted in the flux direction for clarity. In the middle panel, only the CR-free and the CR corrected spectra are plotted to show more detail. In the bottom panel, the relative difference between the CR-free and the CR corrected spectrum is shown.
\label{spsam}}
\end{figure}

For spectroscopic data, the extracted spectra are more
important than the flux of individual pixels on the 2D image.
Figure  \ref{extr} compares the fluxes of the CR polluted part of the
extracted spectra with the CR-free spectra in different
situations: from top to bottom, it shows the falsely detected,
the properly detected, and the undetected CRs, respectively; in
all cases, the average difference between the CR corrected
spectra and the original spectra is less than 2.2\%, as shown
from the distribution in the right column of Figure  \ref{extr} .
Figure  \ref{spsam} shows an extracted spectrum sample; the residual of the CR
correction is within a few percent.

\subsection {Real Data}
\begin{figure}
\includegraphics[scale=.48]{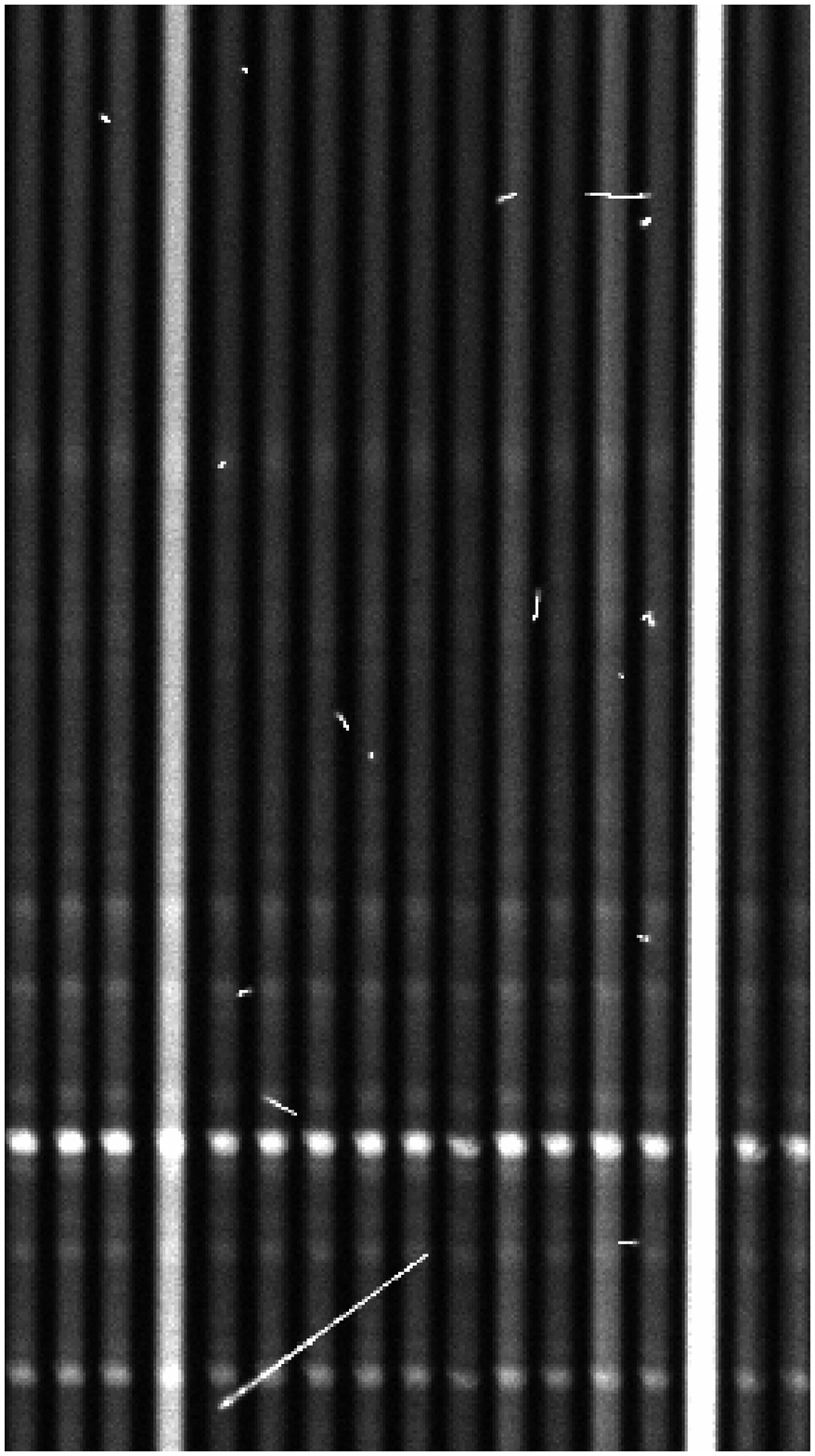}
\includegraphics[scale=.48]{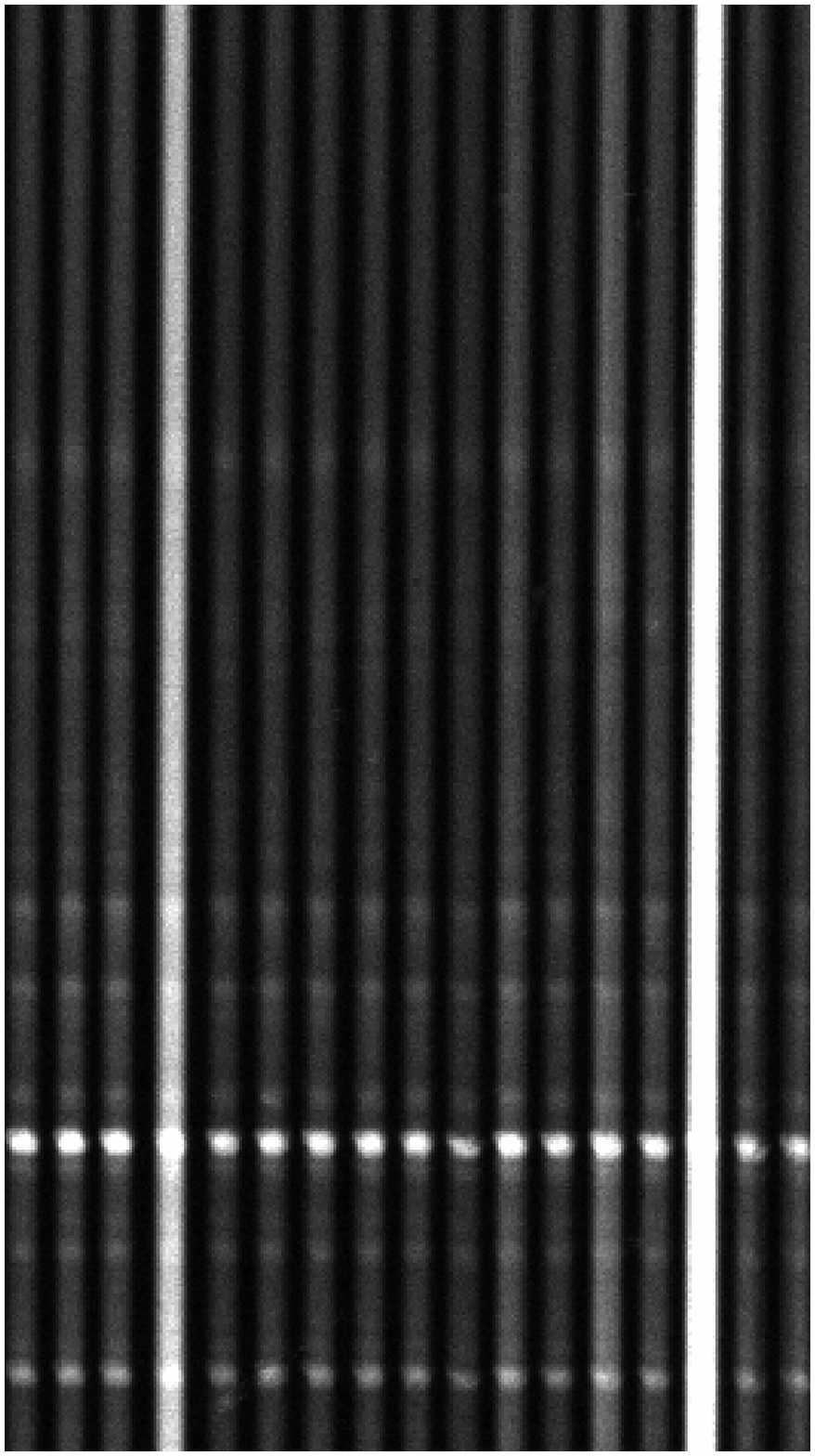}\\
\caption{Left panel is part of a real image from the LAMOST survey, the right panel shows the CR corrected image by our algorithm.
\label{realdata}}
\end{figure}

\begin{figure}
\epsscale{.99}\plotone{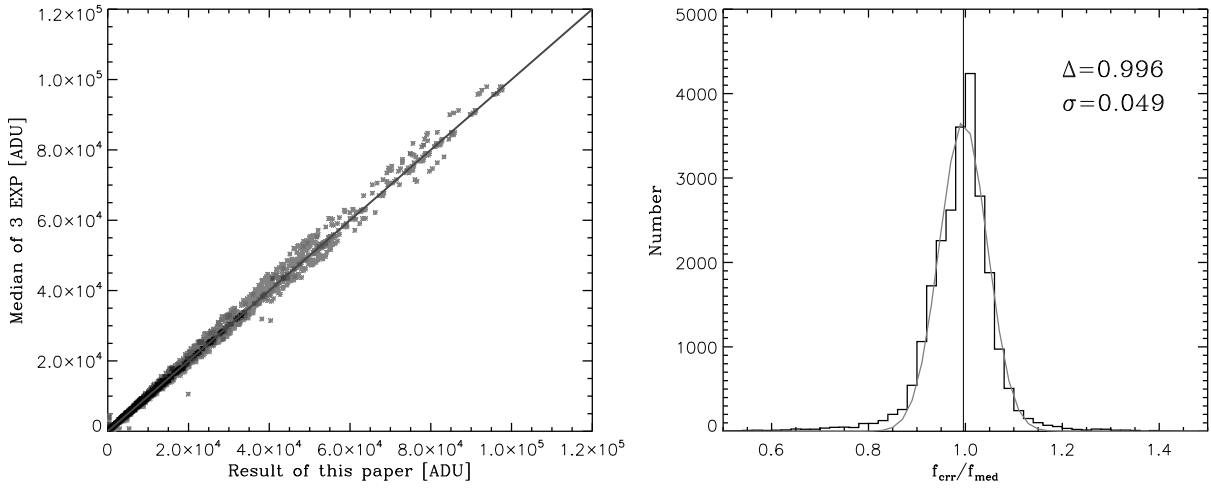}
\caption{Replacement performance on extracted spectra of real LAMOST data. In the left panel, the horizontal axis represents the spectral flux extracted from the CR corrected image and at the position where there is any detected CR within the extraction aperture($f_{crr}$); the vertical axis is the corresponding flux from the image combined from 3 consecutive exposures ($f_{med}$). Right panel: histogram of $f_{crr}/f_{med}$; the parameters of the Gaussian fit of the histogram are marked in the panel. Compared with Figure \ref{extr}, the scatter is larger because the uncertainty in the real data is larger than simulations.
\label{ledfop2com3}}
\end{figure}

We test our algorithm with real data from the LAMOST survey.
Figure  \ref{realdata} shows a part of an 1800s LAMOST image
(left panel) and its reconstruction by our method (right panel).
Visual inspection of our reconstructed image shows that most
of the CR hits are properly removed. For a further comparison,
the extracted flux of the CR detected pixels vs the flux of the
classical multi-exposure-combination method are demonstrated
in Figure ~\ref{ledfop2com3}.
The results are comparable to the simulations in Figure 
\ref{extr}, except that the scatter is a bit larger. The reasons for
the larger scatter could be the following: first, the simulated
image in Figure  \ref{extr} has a higher SNR than the real data; second,
the falsely detected CRs cannot be discriminated from the true
CRs in this test, so the scatter should be larger than the
true-CR-only situation; and third, the sky flux varies between
exposures and the object flux gathered by LAMOST varies
with telescope pointing, which make the combined image
deviate from individual exposures, leading to a larger scatter.

\section {Conclusion}
We present a method for detecting and removing 2D profile
fitting to each segment. A new cosmic ray list is generated by
comparing the fitting residual with a noise model depending on
both the intrinsic shot noise and the relative position in the
profile. We finally produce a more accurate cosmic ray mask
table and more reasonable substitution values for CR polluted
pixels. The method is tested by both simulations and real data;
the results show that our method has a high detection rate, low
false detection rate, and proper replacement of the CR polluted
pixels.

Since this method fits the 2D profiles of the fiber spectro-
scopic data, which are different from the photometric PSF, it
cannot be applied to photometric data. However, it can be used
in slit spectroscopic data after minor modifications. The code
and samples are available at
\url{http://lamostss.bao.ac.cn/~bai/crr}.

Z. Bai acknowledges the support of the National Natural
Science Foundation of China (NSFC) (grant no. 11503054).
H. Zhang acknowledges the support of NSFC Key Program
(grant no. 11333004) and the National Key Basic Research
Program of China (grant 2014CB845700). The Guoshoujing
Telescope (the Large Sky Area Multi-Object Fiber Spectro-
scopic Telescope, LAMOST) is a National Major Scientific
Project built by the Chinese Academy of Sciences. Funding for
the project has been provided by the National Development and
Reform Commission. LAMOST is operated and managed by
the National Astronomical Observatories, Chinese Academy of
Sciences.

\clearpage

\clearpage

\clearpage

\clearpage

\clearpage

\clearpage

\clearpage

\clearpage

\clearpage

\clearpage
\begin{table}
\begin{center}
\caption{Results of CR Detection on Simulated Images \label{table1}}
\begin{tabular}{c|cc}
\tableline\tableline
Items & IMG600 & IMG1800\\
\tableline
CR added & 227451 & 227451\\
CRs detected by this paper & 167844 & 184019\\
CRs detected by L.A.C\scriptsize{OSMIC}\normalsize{ } & 163434 & 173963\\
efficiency of this paper & 73.8\% & 80.9\% \\
efficiency of L.A.C\scriptsize{OSMIC}\normalsize{ } & 71.9\% & 76.4\% \\
False detections of this paper & 5820 & 16626\\
False detections of L.A.C\scriptsize{OSMIC}\normalsize{ } & 559414 & 38912\\
\tableline
\end{tabular}
\end{center}
\end{table}

\end{document}